\def\year{2021}\relax
\documentclass[letterpaper]{article} 
\usepackage{aaai21}  
\usepackage{times}  
\usepackage{helvet} 
\usepackage{courier}  
\usepackage[hyphens]{url}  
\usepackage{graphicx} 
\usepackage{natbib}  
\usepackage{caption} 
\usepackage{xspace}
\usepackage{hyperref}
\usepackage{caption}
\usepackage{cleveref}
\usepackage{subcaption}
\usepackage{booktabs}
\usepackage{float}
\usepackage{xcolor}
\usepackage{cleveref}
\usepackage{xspace}
\newcount\Comments  

\newcommand{\kibitz}[2]{\ifnum\Comments=1{\color{#1}{#2}}\fi}

\newcommand{\condone}{\emph{the Control condition}\xspace}
\newcommand{\condtwo}{\emph{the Label condition}\xspace}
\newcommand{\condthree}{\emph{the Explanation Label condition}\xspace}

\newcount\CommentsNew  
\newcommand{\kibitzNew}[2]{\ifnum\CommentsNew=1{\color{#1}{#2}}\fi}

\title{Do explanations increase the effectiveness of AI-crowd generated fake news warnings? }
\author{            
    Ziv Epstein,\textsuperscript{\rm 1*}
    Nicol\`{o} Foppiani,\textsuperscript{\rm 2*}
    Sophie Hilgard,\textsuperscript{\rm 3*}
    Sanjana Sharma,\textsuperscript{\rm 4*}
    Elena Glassman, \textsuperscript{\rm 3}
    David Rand\textsuperscript{\rm 5}
    \\
}
\affiliations{

    \textsuperscript{\rm 1} MIT Media Lab,
    \textsuperscript{\rm 2} Physics Department, Harvard University
    \textsuperscript{\rm 3} Computer Science Department, Harvard University
    \textsuperscript{\rm 4} Graduate School of Design, Harvard University
    \textsuperscript{\rm 5} MIT Sloan School\\
    \textsuperscript{\rm *} These authors contributed equally

}

\begin{document}

\maketitle

\begin{abstract}
 Social media platforms are increasingly deploying complex interventions to help users detect false news. Labeling false news using techniques that combine crowd-sourcing with artificial intelligence (AI) offers a promising way to inform users about potentially low-quality information without censoring content, but also can be hard for users to understand. In this study, we examine how users respond in their sharing intentions to information they are provided about a hypothetical human-AI hybrid system. We ask i) if these warnings increase discernment in social media sharing intentions and ii) if explaining how the labeling system works can boost the effectiveness of the warnings. To do so, we conduct a study ($N=1473$ Americans) in which participants indicated their likelihood of sharing content. Participants were randomly assigned to a control, a treatment where false content was labeled, or a treatment where the warning labels came with an explanation of how they were generated. We find clear evidence that both treatments increase sharing discernment, and directional evidence that explanations increase the warnings' effectiveness. Interestingly, we do not find that the explanations increase self-reported trust in the warning labels, although we do find some evidence that participants found the warnings with the explanations to be more informative. Together, these results have important implications for designing and deploying transparent misinformation warning labels, and AI-mediated systems more broadly.
\end{abstract}

\section{Introduction}

In recent years, there has been increased societal attention and governmental scrutiny on the spread of misinformation on social media platforms \cite{vosoughi2018spread}, including concerns around political manipulation \cite{roose_2018} and public health \cite{bagherpour_2020} (e.g., the COVID-19 pandemic).
In an effort to curb the spread of misinformation while avoiding direct censorship or content moderation, platforms such as Facebook and Twitter have adopted policies that attach warning labels to certain posts, indicating that they may contain false or disputed information \cite{roth2020updating, constine2016facebook, mosseri2017working}. Previous work has shown that labeling can reduce the rate at which users believe \cite{pennycook2018prior, pennycook2020implied, clayton2019real} and share \cite{yaqub2020effects, pennycook2018prior, pennycook2020implied} misinformation. 



However, these studies also highlight the enormous design space of such warning labels, whereby effectiveness could depend on factors such as the content and source of the label \cite{vraga2017using}, the political affiliation of the user \cite{pennycook2019fighting}, the presence or absence of warning labels on other items \cite{pennycook2020implied}, label humorousness \cite{garrett2019flagging}, and the signal word \cite{sherman2020designing}.

Results from explainable machine learning \cite{lai2019human, bansal2020does} suggest that adding explanations to a computer recommendation may increase the rate at which people accept that recommendation.  While explanations have been shown to increase the trust and efficacy of human-AI systems in other domains \cite{lai2019human, ribeiro2016should} (particularly when the recommendations arise from a complex, algorithmic process \cite{rader2018explanations}), it is unclear how they will work in the context of social media, where attention is scarce,  users are distracted by design \cite{pennycook2021shifting}, and the key outcome is the effect on actual sharing and engagement behavior. 
In this work, we therefore ask whether explaining a system that generates
misinformation labels may similarly increase user sharing discernment (the difference in sharing rates of true and false news items). To decouple the perceptions of warning labels with the particular accuracy of any given labeling system, we focus on a hypothetical system that is 100\% accurate, i.e., only and all known misinformation receives a label. 
In particular, we study the role of explanations on warning labels generated by a (simulated) hybrid human-AI misinformation labeling system.

We focus on a hybrid AI-crowd  labeling system 
because it is highly practical relative to several alternatives.  Many currently deployed misinformation labeling solutions rely on matching claims in social media posts to those disputed by professional fact checkers \cite{lyons_2018}.  However, these expert-based solutions cannot keep up with the scale and speed of misinformation, since misinformation can be produced faster than it can be disproven. 
Indeed, even among posts regarding known debunked claims, labeling can be inconsistent \cite{zannettou2021won}. 
Due to these inherent delays, much violating content might never be labeled with a warning, and even those that do get labeled will only be after the period of maximal exposure. 
This inability of expert-based solutions to keep up with the volume of misinformation may also inadvertently encourage belief in misinformation through the ``implied truth effect,'' whereby users may perceive the lack of a warning label as validation that a piece of content has been fact checked \cite{pennycook2020implied}. In addition, methods employing professional fact-checkers suffer from a general lack of bi-partisan trust: one study from the Pew Research Center found that 70\% of Republicans and 48\% of Americans say fact-checking efforts tend to ``favor one side'' \cite{walker2019republicans}.
Another proposed alternative is fully algorithmic methods for detecting misinformation \cite{bozarth2020toward, abdali2021identifying, weinzierl2021misinformation} (see Related Work for a review). However, these methods fundamentally struggle to keep up with the the non-stationarity of misinformation, and require ground-truth labeling.  Hybrid AI-crowd systems represent a promising alternative that avoids the scalability problems of professional fact-checking by using non-expert human raters to crowdsource the evaluation of information accuracy \cite{kim2018leveraging, pennycook2019fighting, allen2020scaling}. These crowd ratings can then be used as inputs for machine learning algorithms, thereby combining the accuracy of human labels with the scale of automated systems.  Platforms have already started to adopt such techniques, e.g. Facebook's Community Review \cite{silverman19} or Twitter's Birdwatch \cite{coleman_2021, flockers}.




While such a system holds much promise, its complexity may undermine its effectiveness. \citet{yaqub2020effects} suggest that warnings attributed to machine learning-based methods may be \textit{less effective} than warnings attributed to human authorities in leading users to increase the quality of the news they share. They find that the warning effectiveness correlates with users' self-reported ``importance'' of the sources, a measure they associate with trust in a given source, among other factors. 
However, the increased efficacy of fact checker and news media labels may also be at least partially attributable to a greater public familiarity with---and therefore greater perceived accuracy of---these label sources.  
Notably, \citet{seo2019trust} fail to replicate effectiveness of a fact-checker based warning with less than perfect accuracy.

With this in mind, we use the context of a hybrid crowdsourcing-machine learning system as we seek to understand the effectiveness of explaining warning labels in reducing intentions to share misinformation. Our inclusion of a baseline condition in which warning labels appear without explanation additionally serves to help us understand users' acceptance of such a system more generally. As \citet{seo2019trust} find explanations of a label attributed to machine learning increase user discernment without increasing self-reported user trust, we additionally investigate whether our effect is correlated with self-reported measures of trust.  Concretely, we seek to further understand:


\begin{enumerate}
\item  What is the baseline effectiveness of warning labels attributed to a hybrid crowdsourcing-machine learning system in reducing user intentions to share false news? Based on prior work \cite{seo2019trust, yaqub2020effects}, we hypothesize the hybrid warning label will also be effective. 
\item Is it possible to increase the effectiveness of labels attributed to a hybrid crowdsourcing-machine learning system by providing a detailed explanation of the system? Based on prior work \cite{seo2019trust, horne2019rating}, we hypothesize explanations may increase sharing discernment.
\item Is increasing trust in the labeling system a necessary intermediary to increasing effectiveness of warning labels with respect to intention to share? We hypothesize that trust is not necessary to improve discernment \cite{seo2019trust}.
\item Do other moderators that have been investigated in prior work on misinformation interventions due to their connection to the psychology of fake news \cite{epstein2021developing, pennycook2021psychology, pennycook2021reducing} impact the effectiveness of providing a detailed explanation of the warning system? 
\end{enumerate}

We find clear evidence that warnings attributed to human-AI labeling systems increase sharing discernment relative to a no-warning control group. 
We also find some evidence that explaining the process of how this system works increases discernment above and beyond the warning itself. 
Furthermore, we find no difference in the measures of trust between the warning labels with and without explanation, suggesting that enhanced user trust is not the mechanism by which explanations may increase discernment.
However, we find that certain moderators, such as the score in a cognitive reflection task and the age of the subjects, increase the efficacy of the explanations.

\section{Related Work}

\subsection{Explanations and Transparency}

The rise of interpretable or explainable machine learning is motivated, among other factors, by the idea that transparency in machine learning systems can build trust in their outputs \cite{lipton2018mythos,doshi2017towards}. Research has shown that even nonsensical explanations \cite{lai2019human} or explanations of incorrect predictions \cite{bansal2020does} can increase human acceptance of machine learning algorithms. Work on human explanations has shown that the effectiveness of uninformative explanations may be due to inattentiveness, as raising the stakes of a decision eliminates the effectiveness of uninformative explanations, while retaining an effect of informative explanations \cite{langer1978mindlessness}.

\citet{kizilcec2016much} studied the effect of varying levels of explanation detail on trust in an algorithmic grading system and found that a medium level of transparency---in which an explanation of the grading \emph{procedure} was provided---tended to increase trust beyond the low transparency condition, while providing specific details of the calculation (high transparency) decreased trust. Kizilcec further finds that the decrease in trust is correlated with a violation of expectations in the output of the algorithm.
\citet{kulesza2013too} used a music recommendation environment to study the effects of varying the soundness and completeness of explanations on users' trust in the system. They find that high soundness and completeness together result in high trust in the system, but that users prefer concrete feature-based explanations to explanations of the algorithmic procedure.

\citet{rader2018explanations} explore the effects of different types of explanations (\emph{What, How, Why} and \emph{Objective}) on users' perceptions of the Facebook news feed algorithm. They find that while these explanations generally increase knowledge and awareness of the algorithmic system, they may decrease user assessments of the system's correctness, fairness, and responsiveness to user controls. \citet{liao2020questioning} developed an Explainable AI question bank to explore the distinct affordances of different types of explanations. They found that how explanations where helpful when users were in a ``quality-control role.''

Explanation depth is also studied in the context of debunking low quality information. \citet{chan2017debunking} present a meta-analysis of debunking studies, in which detailed debunking messages are found to be more effective than less detailed messages. The authors hypothesize, in line with previous works, that a debunking message must be sufficiently detailed to provide users with an alternative account of the false phenomenon. However, \citet{lewandowsky2012misinformation} recommend using the simplest messages that satisfy this property. In line with this, \citet{martel2021you} show that correction style (including simple versus deep explanations) has minimal impact on correction efficacy. 

\begin{figure*}[h]
    \centering
    \includegraphics[width=0.99\textwidth]{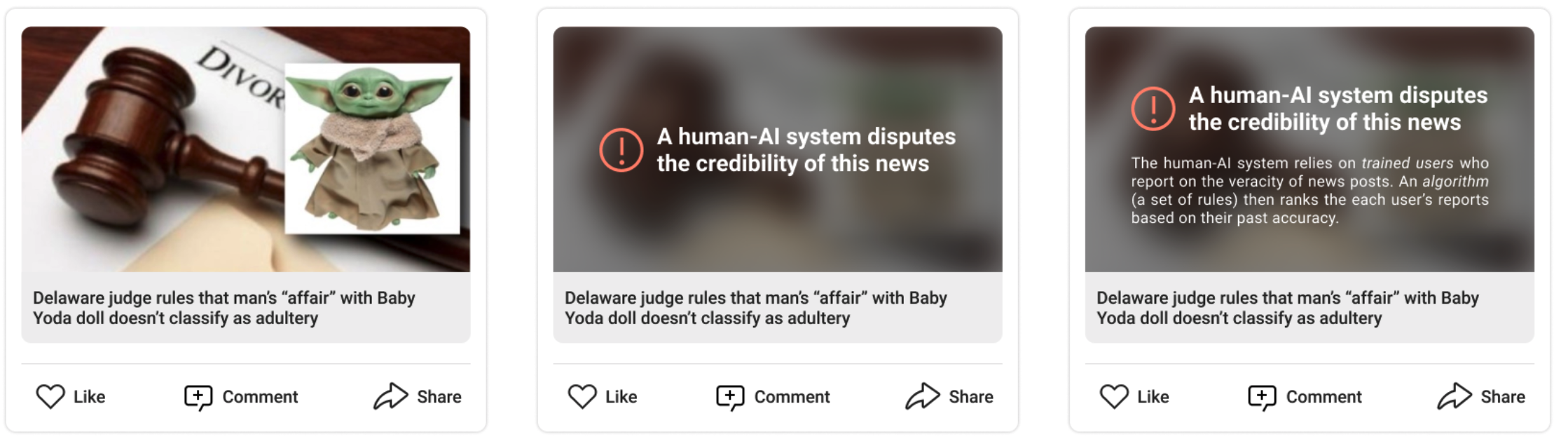}
    \caption{Example of a false post across the three study conditions: \condone (far left), \condtwo  (center), and \condthree (far right).}
    \label{fig:stim_examples}
\end{figure*}
In the context of AI-generated flags for misinformation on social media, \citet{horne2019rating} compared warning labels with explanations to those without in an experimental design similar to ours. They however focus on article reliability as a key outcome, while we focus on sharing. They find that warning labels generated with AI improve perceptions of reliability, and find heterogeneity among participants' baseline perceptions of reliability. 


\subsection{Approaches to Detecting Misinformation}
In an attempt to detect misinformation at a scale necessary for social media, several approaches have been proposed. One class involves computational methods to detect misinformation \cite{bozarth2020toward}.  These computer-based approaches have shown remarkable success, leveraging signals such as sharing patterns \cite{rosenfeld2020kernel}, text features \cite{granik2017fake}, account activity \cite{breuer_eilat_weinsberg_2020}, user stance \cite{weinzierl2021misinformation} and visual features for website screenshots \cite{abdali2021identifying}.  However, the nuanced nature of truth, the limited availability of labeled training data \cite{rubin2015deception, bozarth2020higher}, and the non-stationarity problem whereby the signatures of misinformation can change rapidly (e.g. with the rise of COVID-19 misinformation) place fundamental limits of the effectiveness of fully automated systems. 

A promising alternative that avoids these concerns involves crowdsourcing the fact-checking process. Indeed, crowdsourced fact checking could demote platforms as centralized authorities of truth, and keep up with the fast pace of new forms of misinformation.
\citet{kim2018leveraging} develop an algorithm for deciding which crowd-flagged items to send to professional fact-checkers for verification, with the goal of minimizing overall exposure to false news. Using datasets from Weibo and Twitter, they demonstrate that this method can be effective in reducing the spread of misinformation. 

\citet{pennycook2019fighting} show that laypeople can successfully distinguish high quality media outlets from those which produce hyper-partisan and blatantly false news. They also show layperson trust judgments of websites are highly correlated with those of professional fact-checkers and propose a modification to the newsfeed algorithm, whereby content from websites which receive low crowdsourced trust judgments is downranked. \citet{epstein2020will} build on this work by showing that when informed that their judgments will be used to fight misinformation on social media, participants still correctly distinguish between high and low quality outlets. This suggests that users would not game such a crowdsourcing mechanism to advance their partisan agendas. 
\citet{allen2020scaling} extend this work to the evaluation of headlines, instead of websites. They show that the ratings of small politically-balanced crowds are equally correlated with fact-checker ratings as the fact-checkers' ratings are correlated with one another.

\subsection{Labeling Misnformation and Credibility Indicators}
The research area of misinformation labeling is rapidly evolving, as academic and industry researchers are eager to understand the full implications of design decisions being made on social media platforms \cite{morrow2020emerging}. \citet{clayton2019real} investigate the effects of the locality and severity of misinformation warnings. They evaluate conditions including a general misinformation warning and advice on how to spot misinformation, with or without specific misinformation labels appealing to fact checkers. Further, fact checker labels vary between describing headlines as ``rated false" and ``disputed." They find that specific labels are more effective than general warnings at reducing the perceived accuracy of false news, and that general warnings have a spillover effect, also reducing the perceived accuracy of true news. In the Appendix, the authors additionally state results for intention to share, finding that users in the labeled conditions are no less likely to share items labeled as ``rated false" compared to unlabeled items.


\citet{pennycook2018prior} find that prior exposure to false news, even when labeled as ``Disputed by 3rd Party Fact-Checkers," increases perceived accuracy of those news items when later viewed without labels. While the primary outcome of interest was perceived accuracy, the authors identify in a secondary analysis that disputed labels decrease users' intentions to share false news items. \citet{pennycook2020implied} study both perceived accuracy and sharing intentions in the presence of labels and find that the presence of ``disputed" or ``false" labels decreases both accuracy and sharing for labeled false items, but reverses  for unlabeled false items, via the  ``implied truth effect." The authors find that this effect can be mitigated by explicitly labeling some verified items as ``true."

While early research in this field often focused on accuracy perceptions as outcome \cite{clayton2019real, pennycook2018prior,bode2015related, horne2019rating}, recent work has shown that accuracy perceptions are not a reliable proxy for intention to share \cite{pennycook2021shifting}, and thus addressing the goal of reducing the spread of misinformation requires measuring the latter. 

\citet{yaqub2020effects} explore the effects of appealing to different authorities in social media misinformation labels and find that users are most likely to change their sharing behavior in response to warning labels citing ``fact checkers'' or ``news media'' as the source of the dispute. A condition citing ``AI'' was among the least effective at reducing misinformation sharing in the study. However, this may be due to the fact that currently deployed misinformation labels generally appeal to fact checkers, and so users currently have a better understanding of the accuracy they can expect from fact checkers. It is difficult to know how users with no past experience would have assigned importance to equally accurate label sources.  
\citet{seo2019trust} similarly find some evidence that users are more likely to correctly detect false news when it is identified by ``fact checkers" rather than ``machine learning." However, this result did not replicate in a follow up study in which they removed news sources. This follow up study also tested the effect of adding more (fabricated) detail to the ``machine learning" warning label and found that users were better able to detect false news in this condition but displayed less self-reported trust in the system, which may be due to the explanations being incongruous with the headlines. 

\section{Methods}
\begin{table*}[h]
\centering
  \caption{Bayesian mixed effects model predicting sharing intent. 
  P (bayes) is the number of posterior samples greater than (less than) zero.
  Rhat measures the potential scale reduction factor (PSRF, \cite{brooks1998general}), which captures the variation within a given chain relative to the variation across chains (at convergence, Rhat = 1).}
  \label{tab:main}
  \begin{tabular}{lcccl}
    \toprule
     &Estimate & Standard Error & p (bayes) & Rhat\\
    \midrule
    {veracity} & 0.267&0.106  &0.008$^{**}$&1.002    \\
    {Label} & -0.155&  0.094 &0.052$^{.}$ &1.007  \\
    {Explanation Label} & -0.161& 0.048 &0.031$^{*}$ &1.003 \\
    {veracity $\times$ Label} &0.135 &  0.064 & 0.019$^{*}$& 1.003  \\
    {veracity $\times$ Explanation Label} &    0.217   &  0 &$<0.001^{***}$ &1.002\\
    {Intercept} & 3.201& 0.099&0$^{***}$&1.003 \\
    \midrule
          { N=29,460} &  & & & \\
    \bottomrule
  \end{tabular}
\end{table*}

\begin{figure*}[h]
    \centering
    \includegraphics[width=0.99\textwidth]{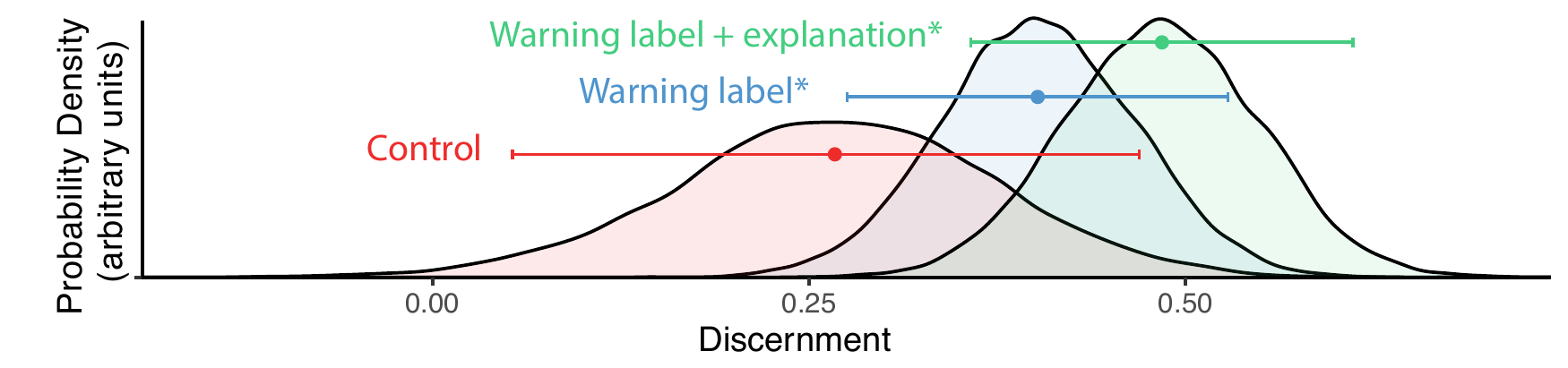}
    \caption{Posterior distributions of the main effect of veracity (e.g. discernment in control, red), and the interactions between veracity and \condtwo and \condthree, plotted using kernel density estimations. 
    The blue and green distributions correspond to the increase in discernment above and beyond the control for the \condtwo and \condthree, respectively. 
    The coloured points show the average values of the distributions, while the error bars illustrate 95\% credible intervals.}
    \label{fig:posterior_distribution}
\end{figure*}
We recruited $N=2512$ Americans using Lucid, which uses quota matching to provide a sample that is nationally representative on age, gender, ethnicity and geographic region \cite{coppock2019validating}. $N=1855$ of these individuals passed two initial trivial quality checks at the start of the survey and were allowed to proceed. Additionally, we only considered subjects who self-reported using at least one social network among Facebook, Twitter, and Instagram. Of these participants, $N=1473$ completed the discernment task, and $N=1394$ completed the entire survey. Unless otherwise specified, we will consider the $N=1473$ subjects who completed the discernment task as our final sample. This final sample had a mean age of 47.87, was 54.1\% female, and was 87\% White. Participants were shown 20 social media  ``cards’’ during the main discernment task, half with a false headline,  half with a true headline. Each participant saw the 20 cards in a random order, and rated their sharing intentions ("How likely would you be to share this story on social media?") on a Likert scale from 1 (Extremely Unlikely) to 6 (Extremely Likely). These headlines were selected from a larger set of headlines \cite{pennycook2020practical} to be relevant to news when the study ran, (e.g. ``stale'' headlines were pulled). 
The stimuli can be found online at \url{https://osf.io/tgwna/?view\_only=efa0ae9ebb30408ca54204e0dc3f53b6}.

In a between subjects design, participants were assigned to one of three conditions: i) a control condition with no labeling (\condone); ii) a basic labeling condition in which users are shown a statement about how the labeling system works prior to any sharing decisions and then presented with the 20 headlines in which false headlines contain a simple label indicating that the claim has been labeled as false by the human-AI system (\condtwo); and iii) an explanation condition which contains all elements of the basic flagging conditions and augments each label with a general (static) explanation of the process by which that label was generated (\condthree).  
See \Cref{fig:stim_examples} for examples of the false items across conditions. 

After the discernment task, all subjects were asked two factual manipulation checks about what they had seen in the images used during the main task \cite{kane2019no}. These questions ensure the treatment was deployed successfully. The subjects who received warning labels then answered a battery of questions related to their opinions of and preferences for the labeling process. Subjects answered a series of multiple choice questions related to how helpful, annoying, and informative these labels are, as well as how often the user would like to see them on social media. 
Finally, they answered a battery of six trust questions derived from \citet{mayer1995trust}'s three factors of trustworthiness: ability, benevolence and integrity (ABI).
For these six ABI questions, we find they are highly correlated ($\alpha$= 0.821), so following standard practice we use Principal Component analysis to construct a single measure of trust that explains 65.3\% of the overall variance.

Users were then asked additional measures, such as two measures of attention \cite{berinsky_attention_checks}, a 6-question cognitive reflection test (CRT) that measures the tendency to stop and think versus going with your gut \cite{crt_Frederick, thomson2016investigating}, political affiliation, and basic demographics. These covariates where chosen because prior work has looked at how they moderate discernment \cite{epstein2021developing, pennycook2021reducing, pennycook2021psychology}, so here we investigate if they impact the effectiveness of the explanation.  Reaction times were not recorded. 

For our main analysis, we fit a Bayesian mixed effects linear model predicting sharing rate as a function of condition and headline veracity and their interaction. Our key  tests are on the interactions between veracity and the treatment dummies (testing whether each treatment increases discernment -- the difference in sharing of true relative to false news -- relative to control). The data is analyzed at the response level, while mixed effects account for variability across subjects and headlines. We adopt a Bayesian modelling approach because the multilevel models with our mixed-effects structure did not converge. We perform full Bayesian inference on the model parameters using the R package BRMS \cite{brms}, which relies on Stan \cite{stan} to sample the posterior distribution using Markov Chain Monte Carlo. We assigned vague and weakly-informative priors, which allow the data to ''speak for themselves” (the model specifications are outlined in the Section 3 of the Appendix). The significance of the coefficients extracted from the data is used as a metric to quantify the observation of an effect. For a given coefficient, we quantify significance with a ``Bayesian p-value'' (pbayes) by counting the percent of posterior samples that are below (or above) zero. The experimental design and analysis methods were preregistered before collecting the data, and are available anonymized at \url{https://aspredicted.org/blind.php?x=2s8eh9}.

\section{Results}
The results of our main analysis are shown in \Cref{tab:main}.  We observe a significant interaction between \condtwo and veracity, indicating that even the simple warning label of the human-AI labeling increases discernment (the difference in sharing of true and false news) relative to the control (pbayes = 0.019, preregistered).  We also observe a significant interaction between \condthree and veracity, such that explaining the label also increases sharing discernment (pbayes $<$ 0.001, preregistered). 
\begin{figure*}[h]
    \centering
    \includegraphics[width=0.99\textwidth]{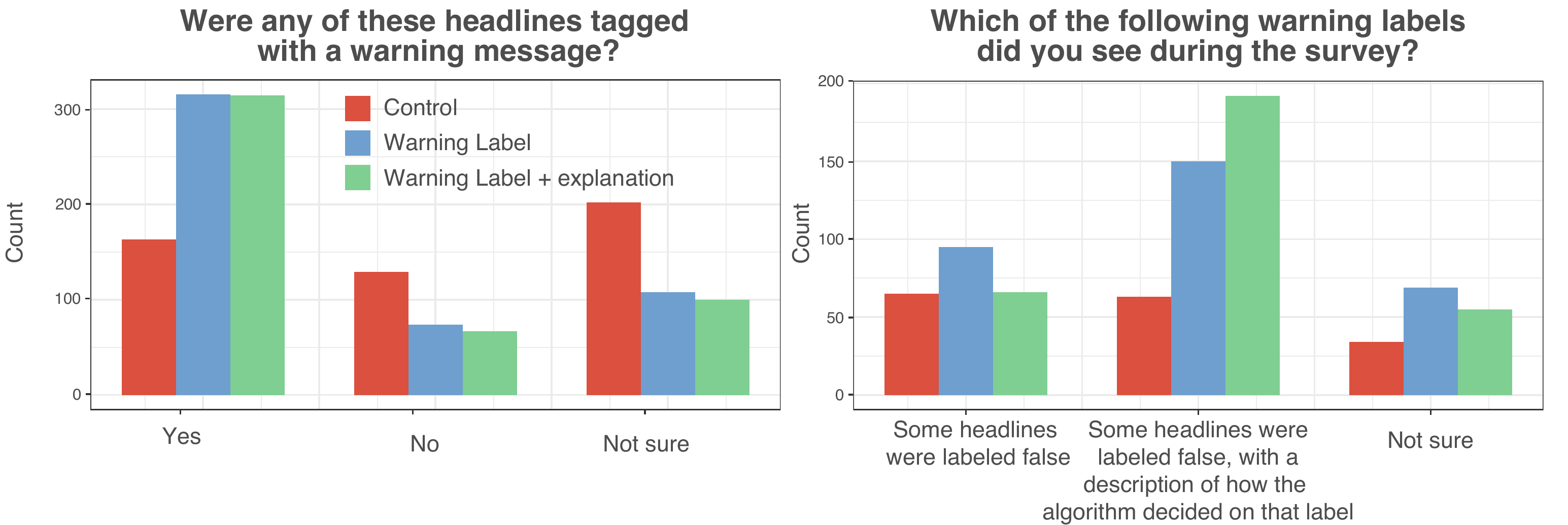}
    \caption{Factual manipulation check for participants in the two treatment condtions.}
    \label{fig:fmi}
\end{figure*}

This main result is summarised in \Cref{fig:posterior_distribution}, where the marginal posterior distributions for baseline discernment (i.e., in control) is plotted, together with the two marginal posterior distributions representing the increase in discernment for the warning label and the warning label with explanation conditions, respectively.

As can be seen from the two interaction coefficients in \Cref{tab:main}, adding the explanation roughly doubled the effect of the warnings on sharing discernment. To assess whether this increase from explanations above and beyond the minimal warning label was statistically significant, we compute the posterior distribution of the difference between the two interaction terms. The mean value is positive, reflecting the observed increase of discernment, but not at a significant level (pbayes = 0.110, pre-registered). This result suggests that an explanation may increase the efficacy of the label above and beyond the warning label itself, but a conclusive answer would require an additional, larger study.

\subsection{The role of trust}
Next, we look at how perceived trust varies across conditions. To do so, we perform a linear regression predicting trust, defined as the first principal component of the ABI questions, as a function of the experimental condition (dummy) and controlling for attentiveness. We do not observe any significant difference in trust between \condtwo and \condthree (p=0.3293). In fact, we directionally observe that more attentive subjects in \condthree report lower trust (p=0.09) but the overall moderation of attentiveness is not significant either.

This suggests that, in so much as the explanation may increase the effectiveness of the warnings, trust in the warning label is not the mechanism by which this occurs.  In addition, we find no significant differences between the two \textit{Warning Label} conditions in participant ratings of how helpful the warning label was (p=0.253), how annoying the warning label was (p=0.345), how well the warning label helped the user understand the labeling process (p=0.169) or if the warning labels helped them understand what signals were leveraged (p=0.56). We do find that participants in \condthree said the warning labels provided new information marginally more than in \condtwo (p=0.067). This may suggest that the explanation provides new information into the process that generated the the labels that the user can use or act on in their sharing decisions.

\subsection{Factual manipulation check}
After the main discernment task, we included manipulation checks to see what participants thought they saw during the discernment task. These results of these manipulation checks is shown in \Cref{fig:fmi}.  Several subjects in \condone report to have seen a warning label, and some participants in both \condtwo and \condthree stated they did not (see the left pane of \Cref{fig:fmi}). 
Participants who said they saw a headline tagged with a warning message were then asked which warning labels they saw. While participants in \condtwo made up the plurality of respondents who saw ``some headlines were labeled false,'' they were also more likely to say they saw a warning label with an explanation than one without (see the right pane of \Cref{fig:fmi}). 
This suggests that our warning label design worked up to a certain extent, but that even in a survey context --- where people are likely paying more attention than when scrolling though social media feeds --- users do not always realise they are presented with warning labels or explanations, or they might believe they have seen warning labels when they were not presented with them.

\begin{figure*}[h]
    \centering
    \includegraphics[width=0.99\textwidth]{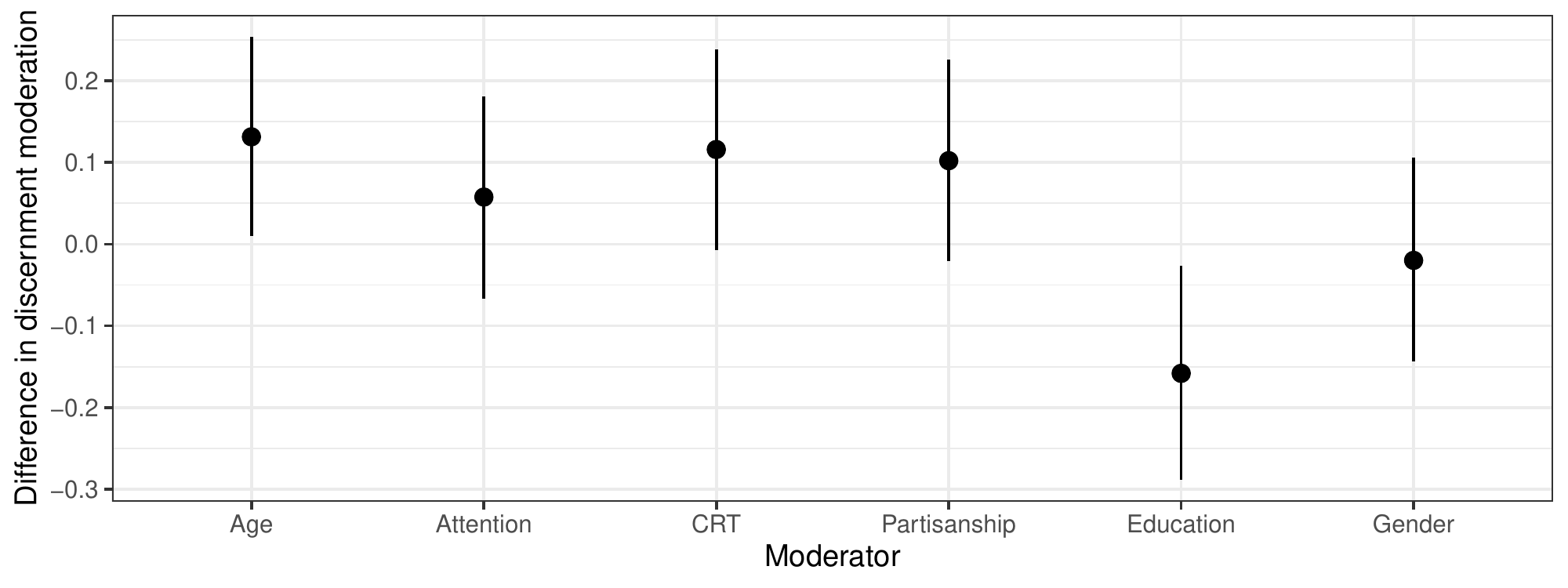}
    \caption{Posterior distributions for the difference in moderation of discernment between the two treatment conditions.}
    \label{fig:moderation}
\end{figure*}

\subsection{Additional moderators}
Additional analyses help us determine which factors moderate the treatment effect. 
We consider attention, propensity to go with one's gut, as measured by the critical reflection test (CRT), partisanship, age, education and gender as possible moderators.  Partisanship is coded as increasing means more Republican.  Gender is coded as 1 as female and 0 otherwise. 

For each of these factors, we modify the model for the main analysis by including this factor as a predictor, and re-run the Bayesian model. Each moderator is scaled by subtracting the average and dividing by the standard deviation. This ensures that the coefficients for the fixed effects are compatible with the main analysis within statistical uncertainties. 


We observe that more attentive participants, higher CRT participants, more conservative participants, older participants, less educated participants, and female participants share less at baseline. More attentive participants, higher CRT participants and older participants are more discerning in their baseline sharing. However, we do not observe any significant interactions involving moderators, except for Education interacting with \condtwo, and attention interacting with \condthree (We also do not account for multiple comparisons here, which would render both interactions non-significant). For statistical details, see Supplemental Table S1. This general lack of moderation suggests that warning labels work equally for a diverse set of users. However, this result fails to give us a sense of the mechanisms by which warning labels increase discernment, which should be explored in future work.  

To better understand the mechanisms behind the increase in discernment due to the explanation itself, we also measure and report the difference in moderation of the treatment effect on discernment between \condthree and \condtwo (see Figure~\ref{fig:moderation}) 
By calculating a credible interval for the difference between the two coefficients, we can assess if explanations are deferentially effective for certain subgroups. We find a significant difference between the 3-way interactions for age, CRT, partisanship and education. This suggests that the explanations are differentially more effective at increasing discernment among participants with lower levels of education or CRT, who are older, and are more conservative. We also observe a directional difference in 3-way interactions for attention, whereby explanations are more effective for attentive users. 


\section{Discussion}
Our results suggest that providing users with warning labels created by a hybrid AI-crowd system can increase sharing discernment relative to a no-label control group. This suggests that labeling with human-AI warning labels has promise. Given the pragmatic merits of this approach for detecting misinformation at scale, our work suggests that people may indeed respond well to such a labeling scheme. 

We also found circumstantial evidence that providing more transparency via explaining what the AI system is doing may boost the system's efficacy above and beyond the basic warning. This adds to the conversation about correction explanations for debunking. Most work on corrective explanations focuses on explanation depth at the headline or claim level~\cite{lewandowsky2021under}, but has not looked at the corrective depth of the debunking \emph{mechanism} itself. Ultimately, this approach is more scalable since all headlines share the same mechanism explanation, and generating headline-level explanations can be costly.  


We also observed several significant moderators: explaining the warning labels was more effective for participants with lower levels of education or CRT, who are older, and are more conservative. These findings have important implications for how future explanations are designed, and at whom they are targeted.  We also found directional moderation of attention, whereby explaining the warning label was more effective for more attentive users. Even in the survey context, where participants are presumably focused on the task at hand, users did not always realize they were presented with warning labels or explanations in the first place, or alternatively they might believe they saw labels when they were not present.  While explanations may provide an important tool for boosting the efficacy of opaque human-AI systems, this effectiveness could be undermined by environments where users are distracted and thus do not even consider them. This finding suggests important avenues for future work. First, it highlights the importance of designing warning labels in general, and explanations in particular, that grab more attention from users. 

Further, our results suggest that (self-reported) user trust in the warning label is not the mechanism by which explanations contribute to the efficacy of machine recommendations. Despite a good deal of research that assumes a causal relationship between interpretability, trust, and acceptance~\cite{ribeiro2016should}, previous works have been equivocal in proving these relationships. \citet{cramer2008effects} find that while explanations increase acceptance of machine recommendations, they do not increase self-reported trust or willingness to delegate to the system on a future task. 
\citet{papenmeier2019model} find that while explanations do not increase self-reported trust, they do increase \emph{observed} trust, as measured by the probability that users switch their pre-reported answers to match those of the machine recommendation.
In their experiment, this does not necessarily increase accuracy, as the algorithm does not achieve 100\% accuracy. Indeed, we note that while achieving greater acceptance of machine recommendations is equivalent to increasing discernment in our experiment, in general the goal of explanations should be to increase \emph{calibrated trust}, that is, trust that is appropriate for the performance of the system \cite{zhang2020effect}. \citet{papenmeier2019model} additionally find that the accuracy of the machine learning system has a greater effect on user trust, both self-reported and observed, than the presence and quality of explanations.

Our results are consistent with \cite{horne2019rating}. While they consider accuracy instead of sharing as the key outcome, our work corroborates the idea that providing explanations to AI labels has potential to augment user discernment. Further, our results support their findings on the heterogeneity of the efficacy of such explanations. 
Our results concerning attentiveness and explanation effect are also consistent with \citet{langer1978mindlessness}. Studying human explanations (rather than AI), the authors found that relevant explanations are in low-stakes problems no more effective than a placebo explanation at increasing user acceptance of a request. However, when the stakes were slightly higher, the quality of the explanation provided did matter. The authors hypothesize that the higher stakes of the request induce attentiveness where respondents were previously inattentive, thereby increasing the importance of quality explanations. We believe this suggests interesting future directions for better understanding the moderators of warning label effectiveness, including evaluating users' trust with more measures than explicit self-reporting and understanding users' pre-existing experience with a given label source (e.g., fact checkers, AI, crowdsourcing) and perceptions concerning the accuracy of that label source.

Our work has several limitations that must be considered. First, our work only considers hypothetical sharing intentions, and therefore may suffer from demand effects. Since previous work has shown that self-reported sharing intentions correlates with actual sharing on Twitter \cite{mosleh2020self} and has replicated misinformation reducing interventions in Twitter field experiments \cite{pennycook2021shifting}, we believe the use of online surveys to be justified by providing a sandbox to test otherwise challenging interventions. Second, in our study only false headlines were labeled by the system. Future work should explore how discernment changes when the system erroneously labels true items. Third, the nature of the human-AI explanation we used, with a vetting process for users and a particular crowd-sourcing methodology, is a specific representation from a large space of possible systems.  We chose this particular process because of its similarity to real world systems, but future work should explore how these results generalize to across other possible systems, such as different kinds of explanations, differently designed labels, whether the AI labels or vets, and expert versus non-expert crowds. Fourth, our study focused on the cultural context of American misinformation, and recruited American participants. Future work should explore how such findings translate to other cultural contexts. In addition, our design cannot disentangle the effectiveness of warning labels attributed to the human-AI system, versus any warning label at all. We chose the human-AI system as a baseline because we are primarily focused on the role of explanations, but more work is necessary to distinguish the effects of more traditional warning labels (e.g. those from \cite{yaqub2020effects}) with the specialized labels from the human-AI system. Future work might do so by including controls for both a standard, non-attributed warning label, and the human-AI baseline we used. Our study also relies on perceptions of a \emph{hypothetical} human-AI labeling system. It is unclear if participants believe that the warning labels actually came from a human-AI system, which may undermine our results. Correspondingly, we did not investigate how the act of sharing intersects the labeled content.  For example, a user might be happy that a piece of content was labeled and might want to share the warning label itself with their friends, which in our data would be interpreted as the warning label not working. These quandaries underscore the challenges of understanding the nuances of genuine social media behavior in a survey context. 

Taking a step back, its important to note that our approach of design interventions to combat misinformation is but one tactic to fight a system with many actors and perverse incentives. Our intervention may systematically benefit large companies by providing them ``band-aid'' solutions to the systemic problems of an attention economy, rather than forcing them to consider the underlying causes. Ultimately, we hope our work invigorates careful research on how conventional approaches to fighting misinformation and making systems transparent intersect with the subtle, complex and consequential psychological dynamics of social media.

\bibliography{refs}

\end{document}